\documentclass[
preprint,
amsmath,amssymb,amsthm,
aps,
floatfix,
]{revtex4-1}

\usepackage[utf8]{inputenc}
\usepackage[english]{babel}
 
\newtheorem{theorem}{Theorem}
\newtheorem{proposition}{Proposition}
\newtheorem{definition}[proposition]{Definition}

\newtheorem{question}[theorem]{Question}
\newtheorem{remark}[proposition]{Remark}

\usepackage{graphicx}
\usepackage{dcolumn}
\usepackage{bm}
\usepackage{xcolor}
\usepackage{psfrag}

\begin{document}


\title{Local minima in disordered mean-field ferromagnets}


\author{Eric Yilun Song}
\affiliation{NYU Shanghai}
\author{Reza Gheissari}%
\affiliation{NYU--Courant Inst.\ of Math.\ Sciences \\ (now at UC Berkeley)}
\author{Charles M.\ Newman}%
\affiliation{NYU--Courant Inst.\ of Math.\ Sciences \\ NYU-ECNU Institute of Mathematical Sciences at NYU Shanghai}
\author{Daniel L.\ Stein}
\affiliation{NYU--Courant Inst.\ of Math.\ Sciences \\ NYU-ECNU Institute of Mathematical Sciences at NYU Shanghai \\ Santa Fe Institute}
%




\date{\today}

\begin{abstract}
We consider the complexity of random ferromagnetic landscapes on the hypercube $\{\pm 1\}^N$ given by Ising models on the complete graph with i.i.d.\ non-negative edge-weights. This includes, in particular, the case of Bernoulli disorder corresponding to the Ising model on a dense random graph $\mathcal G(N,p)$. Previous results had shown that, with high probability as $N\to\infty$, the gradient search (energy-lowering) algorithm, initialized uniformly at random, converges to one of the homogeneous global minima (all-plus or all-minus). Here, we devise two modified algorithms tailored to explore the landscape at near-zero magnetizations (where the effect of the ferromagnetic drift is minimized). With these, we numerically verify the landscape complexity of random ferromagnets, finding a diverging number of (1-spin-flip-stable) local minima as $N\to\infty$. We then investigate some of the properties of these local minima (e.g., typical energy and magnetization) and compare to the situation where the edge-weights are drawn from a heavy-tailed distribution. 
\end{abstract}



\maketitle

\section{Introduction}\label{sec:level1}

One of the distinguishing features of discrete-spin systems with quenched disorder---especially those whose disorder resides in the couplings---is the complexity of their energy landscape (\emph{Hamiltonian}) as the system size $N$ diverges. This complexity can be formalized in many ways but a canonical choice is the presence of many (diverging rapidly as $N\to\infty$) metastable traps. Precise definitions will be given below, but informally a trap is a spin configuration (or a dynamically-connected set of spin configurations) which is not a global minimum of the Hamiltonian but which at zero temperature is absorbing in the dynamics. The precise definition of a trap therefore depends on the dynamical rules governing the time evolution of the system, but should be understood as a \emph{local minimum} of the landscape. 

In this paper, we will consider the widely studied Glauber-type dynamics consisting of asynchronous updating of randomly chosen spins which then flip (or don't flip) according to a prespecified rule. That is, the traps we consider are the usual 1-spin-flip metastable states (or sets of states) as determined by the \emph{local minima} of the Hamiltonian: when Glauber-type dynamics are run at zero-temperature (roughly in analogy with gradient descent in continuous spaces), these 1-spin-flip stable states are the absorbing states of the dynamics.  We will focus on disordered Curie-Weiss ferromagnets, i.e., Ising systems of $N$ $\{\pm 1\}$-spins on a complete graph with edge weights (i.e., couplings) that are nonnegative i.i.d.\ random variables. More precisely, the energy landscapes we consider are of the form 
\begin{align}\label{eq:Hamiltonian}
    \mathcal H(S): = -\sum_{1\leq i<j\leq N} J_{ij} S_i S_j\,, \qquad \mbox{for $S \in \{\pm 1\}^N$}\,,
\end{align}
where $J_{ij}=J_{ji}$ are i.i.d.\ drawn according to some $\nu$ supported on $\mathbb R_+$.

The classical zero-temperature Glauber dynamics for such systems initializes from a configuration $S(0)$ according to some prior distribution on $\{\pm 1\}^N$, then selects an $i \in \{1,..,N\}$ uniformly at random and flips its spin if such a spin flip decreases $\mathcal H$ (see Section~\ref{sec:level2} for a more precise definition).  In~\cite{GNS18}, the dynamics for the \emph{randomly diluted Curie-Weiss model} ($J_{ij}$ are~1 with probability~$p$ and~$0$ with probability~$1-p$) with an initial state chosen uniformly at random, was analyzed. (This initialization models an instantaneous quench from infinite to zero temperature.) It was proven in that paper, that in such a scenario the system absorbs, with probability approaching one as $N\to\infty$, into one of the two uniform spin configurations (all-plus or all-minus); in particular it does not feel any complexity in the landscape $\mathcal H$. It was further argued in~\cite{GNS18} that this  should hold for any light-tailed distribution, such as half-normal; that was confirmed numerically in~\cite{GNSW}, and again here.

This result is particularly interesting because the dynamical behavior just described is~{\it not\/} what is seen in mean-field spin glasses (i.e., the SK model~\cite{SK75}) or  spin glasses and random ferromagnets in any finite dimension~\cite{YGMNS17,NS99c,NS00jl,NNS00}. In these systems,  zero-temperature dynamics is heavily dominated by the presence of numerous traps, and the probability of the dynamics avoiding such traps and ending in one of the two uniform states (or in the case of the spin glass models, one of the global energy minima) goes to zero as $N\to\infty$. The result of~\cite{GNS18} then raises the question whether metastable traps are absent altogether from random Curie-Weiss models, or are they present but in sufficiently small quantity that the measure of the union of their domains of attraction rapidly goes to zero as $N\to\infty$; the latter situation would mean that the number and sizes of traps increase sufficiently slowly with $N$ that a typical dynamical run of zero-temperature dynamics would not find any of them. 

To better understand the structure of the energy landscape, particularly in comparison to mean-field spin glass models which are much more extensively studied, let us decompose the Hamiltonian. Letting $\lambda=\mathbb E[J_{12}]$ be the mean of the distribution $\nu$ so that $\lambda>0$,
\begin{align}
\mathcal H(S) = -\sum_{i<j} J_{ij} S_i S_j & = -\sum_{i<j} [J_{ij} -\lambda] S_i S_j - \lambda \sum_{i<j} S_i S_j \nonumber \\ 
& = - \sum_{i<j} \tilde J_{ij} S_i S_j - \lambda( \sum_i S_i)^2\label{eq:Hamiltonian-splitting}
\end{align} 
where $\tilde J_{ij}$ is mean zero. Although $\tilde J_{ij}$ does not necessarily have a symmetric distribution, as it is non-degenerate and mean zero, the term $\tilde {\mathcal H} (S):= - \sum_{i<j} \tilde J_{ij}S_i S_j$ can be understood as having the properties of a mean-field spin glass landscape (and in the case where $\nu$ is Bernoulli with $p=1/2$, it will indeed be a $\pm J$-spin glass). The second term is a quadratic potential in the magnetization $M(S):= \sum_i S_i$ whose strength depends linearly on the mean of $\nu$. In particular, when $M(S)$ is small, $\mathcal H(S)$ is close to $\tilde {\mathcal H}(S)$, whereas when $M(S)$ is order $N$, $(M(S))^2$ dominates the first term whose typical values are $O(N)$ and extremal values are $O(N^{3/2})$.  

As a consequence, the role played by the quadratic potential is to wash away much of the complexity of the landscape $\tilde {\mathcal H}(S)$ away from $M(S)=0$, by introducing a strong drift towards the fully magnetized all-plus and all-minus configurations. This suggests that our goal of analyzing the remaining complexity of the landscape at  near-zero magnetizations requires us to substantially alter the traditional zero-temperature Glauber dynamics. 

After formalizing the setup of the paper, and defining traps/local minima more precisely in Section~\ref{sec:traps}, we propose in Section~\ref{sec:level2} two different search algorithms whose transition rates are designed to counteract the ferromagnetic drift and search for (non-uniform) local minima of $\mathcal H(S)$. Using these, we are able to establish numerically that with high probability the landscape $\mathcal H(S)$ has a diverging (in $N$) number of local minima; moreover, the time and computational complexity of finding such local minima both appear to be polynomial in $N$. From there, we study more refined features of these local minima such as their typical magnetizations and energies. These numerical results are presented in Section~\ref{sec:results}.

We end the paper in Section~\ref{sec:discussion} with an extended discussion of the relationship between the present problem and extensively studied questions of landscape complexity and algorithmic difficulty in spin glasses, spiked matrix and tensor models, and constrained optimization problems. We pose a series of  questions to motivate future numerical and theoretical analysis of disordered mean-field ferromagnets in that context. 

\section{Landscapes and Traps}
\label{sec:traps}

In this section we formalize the landscapes we consider in this paper and the definition of traps or local minima of these landscapes. We will start with some energy landscape, or \emph{Hamiltonian} $\mathcal H:\{\pm 1\}^N \to \mathbb R$ assigning a real value (corresponding to the configuration energy in physical applications) to each spin configuration on $N$ spins, viewed as an element of the hypercube $S\in \{\pm 1\}^N$. We will also assume a natural 1-spin-flip asynchronous dynamics as described in the Introduction that never increases its energy (i.e., is \emph{zero-temperature}). Our Hamiltonians will be of the form
\begin{equation}
\label{eq:energy}
    \mathcal H(S) = -\sum_{i<j} J_{ij} S_i S_j 
\end{equation}
for i.i.d.\ couplings $(J_{ij})_{i\neq j}$ drawn from a common distribution $\nu$ with non-negative support. The latter constraint ensures that the model is \emph{ferromagnetic} and the ground states (global minima of $\mathcal H(S)$) are the uniform all-plus and all-minus configurations. 

An important case we consider is where $\nu$ is Bernoulli with parameter $p \in (0,1)$, so that 
\begin{equation}
    J_{ij} = \begin{cases}1, & \quad \mbox{with prob.\ $p$}, \\ 0, & \quad \mbox{with prob.\ $1-p$}. \end{cases} 
\end{equation}
This choice of coupling distribution corresponds to the classical Ising model on a dense Erdos--Renyi random graph $\mathcal G(N,p)$ and was also the main object of study in~\cite{GNS18}. 

We will also study some other choices of $\nu$ to understand the universality (or lack thereof) of the structural properties we find. For this purpose, we study one other light-tailed distribution, and one heavy-tailed distribution (numerical studies which we do not include in this paper suggest that the specific choice of distribution among these does not qualitatively change the properties of the landscape).  

Our choice of a light-tailed distribution with density is the half-normal distribution where 
\begin{equation}
    J_{ij}\sim |Z| \quad \mbox{for}\quad Z\sim \mathcal N(0,1)\,.
\end{equation}
Finally, our choice of a heavy-tailed distribution is the half-Cauchy distribution $J_{ij}\sim |Y|$ where $Y$ has density 
\begin{equation}
    \label{eq:Cauchy}
p_Y(y) = (\pi (1+y^2))^{-1}\, .
\end{equation}

\begin{definition}
A \emph{1-spin flip stable state} of the energy landscape $\mathcal H$ is a configuration $S\in \{\pm 1\}^N$ such that for every $S'$ whose Hamming distance $d_H(S, S')=1$, we have $\mathcal H(S')\geq \mathcal H(S)$. 
\end{definition}

In ferromagnetic models, the two uniform (all-plus and all-minus) configurations are always global minima and are therefore also 1-spin flip stable.  
For the purposes of this paper, we will refer to those configurations only as the uniform or homogenous configurations, and let \emph{local minima} refer to all 1-spin flip stable states that are neither of the two uniform configurations. 

\begin{remark}
In models with continuous disorder, e.g., half-normal, almost surely no two neighboring configurations have the same energy. However, in models whose disorder distribution has atoms, like the Bernoulli disorder, it is possible that neighboring configurations have equal energy. In this case, we distinguish \emph{traps} which are finite connected (Hamming distance one) sets of local minima $S_1,S_2,...,S_k$ of the hypercube $\{\pm 1\}^N$; in this case, we say $k$ is the number of members of a trap. 
\end{remark}

Returning to random Curie-Weiss models, we reiterate that Theorem~1 of~\cite{GNS18} implies that for large systems, the traditional greedy algorithms (e.g., zero-temperature Glauber dynamics) described in Sec.~\ref{sec:level2} will absorb in one of the all-plus or all-minus ground states. In fact, it was found that with high probability, the absorbing ground state is dictated by the initial $O(\sqrt N)$ bias in the magnetization $M(S(0))$, and the absorption happens in time that is $O(N\log N)$.  The theorem was proved for the randomly diluted model where the edge-weights are sampled from i.i.d.\ $\mbox{Ber}(p)$ for $p\in (0,1)$. The heuristic driving the proof carries to all light-tailed coupling distributions, and therefore it is very likely that the same conclusions apply to the random Curie-Weiss ferromagnet with i.i.d.\ couplings drawn from any light-tailed distribution  with, say, exponential moments: see also \cite[Conjecture 1]{GNSW}.  

The typical absorption into the all-plus and all-minus states derives from the fact that at configurations with magnetizations that are order $\sqrt N$, the drift from the quadratic potential in~\eqref{eq:Hamiltonian-splitting} is on dominates the drift from fluctuations in the centered (glassy) part of the landscape. 
As such, with high probability, there are no 1-spin flip stable states with magnetizations $M\gg\sqrt N)$ except the uniform all-plus and all-minus states. As such, the complexity of the landscape $\mathcal H$ is confined to magnetizations that are $o(\sqrt N)$, where the landscape is not trivialized by the ferromagnetic drift. Indeed, our search algorithms that explore $\mathcal H$ looking for local minima (described in the following section) are designed to counteract this drift and minimize the absolute magnetization as much as possible.

\section{\label{sec:level2}Markov Chains searching for local minima}
In this section, we define the search algorithms we use to find local minima of random ferromagnets, and discuss their implementation in the numerics contained in this paper. Let us begin by recalling the classical zero-temperature Glauber dynamics chain. The traditional zero-temperature dynamics used to optimize on the landscape $\mathcal H$ is known as the \emph{zero-temperature Glauber dynamics} (seen as a zero-temperature limit of the usual Glauber dynamics~\cite{Gl63}). That dynamics is a Markov chain $(X(t))_{t\in \mathbb N}$ on $\{\pm 1\}^N$ which is initialized from a random configuration,(i.e.,  $X_i(0)$ are i.i.d.\ $\pm 1$ with probability $\frac 12$ each) and for every $t\in \mathbb N$, updates a configuration $X(t)$ to $X(t+1)$ as follows: 
\begin{enumerate}
    \item Select a site $i\in \{1,...,N\}$ uniformly at random.
    \item If $\sum_{j \neq i} J_{ij} X_j(t)=0$, flip a fair $\pm 1$-coin for $X_{i}(t+1)$.
    \item Otherwise, set $X_i(t+1)= \mbox{sign}[\sum_{j \neq i} J_{ij} X_j(t)]$. 
\end{enumerate}
As mentioned earlier, it was proven in~\cite{GNS18} that this zero-temperature search algorithm, initialized from a uniformly random configuration, absorbs into one of the homogeneous ground states with probability $1-o(1)$. Moreover, though not proved, this $o(1)$ appeared to be decaying very rapidly in $N$, perhaps exponentially, and this phenomenon persisted even when the initialization was forced to have magnetization zero (see Figure~\ref{fig:diffalg}).

\subsection{Constrained zero-temperature dynamics}\label{subsec:constrained-alg}

We introduce a \emph{constrained zero-temperature dynamics} designed specifically to avoid ending up in one of the homogeneous ground states, and to increase the probability of ending up in a local minimum of $\mathcal H$. Towards this goal, the constrained dynamics modifies the usual zero-temperature dynamics in several ways which we now describe. A run of the \emph{constrained zero-temperature dynamics with boundary $M_b$} is defined as follows (suppose $N$ is even, for simplicity). Initialize $X(0)$ randomly among configurations with magnetization zero ($\sum_i X_i(0)= 0$) and for every $t\in \mathbb N$, update a configuration $X(t)$ to $X(t+1)$ as follows: 
\begin{enumerate}
    \item Select a site $i\in \{1,...,N\}$ uniformly at random.
    \item If $\sum_{j \neq i} J_{ij} X_j(t)= 0$, let $X(t+1) = X(t)$.
    \item Otherwise, let $X'$ be the configuration $X(t)$ updated to have $X'_i= 
    \mbox{sgn}[\sum_{j \neq i} J_{ij} X_j(t)]$ and if 
    \[ \big|\sum_i X'_i\big| \leq M_b ,
    \]
    set $X(t+1) = X'$. Otherwise, let $X(t+1) = X(t)$. 
\end{enumerate}
The algorithm either terminates in a local minimum or in a boundary configuration with $|\sum_i X_i(T)|=M_b$ whose only 1-spin flip neighbors $S$ with $\mathcal H(S)\leq \mathcal H(X(T))$ have absolute magnetization $|\sum_i S_i|>M_b$. 

The crucial modifications here are the initialization with magnetization zero and the imposition of a boundary constraint on the absolute magnetization; this boundary constraint ensures that the ferromagnetic mean-field effect is minimal and the landscape as seen from the dynamic search ``looks more glassy"; the idea is that if the dynamics is forced to spend a sufficiently long time in the constrained region of magnetizations, it finds a local minimum before the ferromagnetic drift dominates.

\subsection{Biased zero-temperature dynamics} 

An alternative approach to the constrained algorithm described above would be to impose a drift towards zero magnetization to counteract the ferromagnetic drift towards the homogenous plus and minus states. However, one would want to implement such a drift without changing the fact that the algorithm seeks out and is trapped by the local minima of the original Hamiltonian $\mathcal H(S)$ (without any external drift term). In view of this, we propose the following \emph{biased  zero-temperature dynamics}. For a configuration $S$ and an $i\in \{1,...,N\}$, let $S^{(i)}$ be the configuration which agrees with $S$ except on $i$, for which $S_i = -S_i^{(i)}$. From a configuration $X(t)$, generate $X(t+1)$ as follows:
\begin{enumerate}
\item If there exists $i\in \{1,...,N\}$ such that $\mathcal H(X^{(i)}(t)) < \mathcal H(X(t))$ and $|M(X^{(i)}(t))|<|M(X(t))|$, then select such an $i$ uniformly at random and let $X(t+1) = X^{(i)}(t)$. 
\item If there does not exist any such $i$, select $i\in \{1,...,N\}$ uniformly at random and if $\mathcal H(X^{(i)}(t))<\mathcal H(X(t))$ let $X(t+1) = X^{(i)}(t)$. 
\end{enumerate}

This is particularly appealing in theory as it is a \emph{smoother} modification than the hard barrier of the constrained zero-temperature dynamics, and simultaneously counteracts the ferromagnetic drift even at very small magnetizations ($0< |M(S)|<M_b)$ where the constrained dynamics would not feel the effect of the barrier. Moreover, as with the usual zero-temperature dynamics, it always absorbs into a $1$-spin flip stable state of $\mathcal H(S)$ which is either a local minimum or uniform global minimum.

\subsection{Implementation}
Following the above intuitions, we will see that the probability that given dynamical runs of both the constrained zero-temperature and biased zero-temperature search algorithms absorb into local minima are substantially larger than those probabilities for the traditional zero-temperature Glauber dynamics, even if we initialize optimally (at magnetization zero).

\begin{figure}[h]
    \includegraphics[scale=0.9]{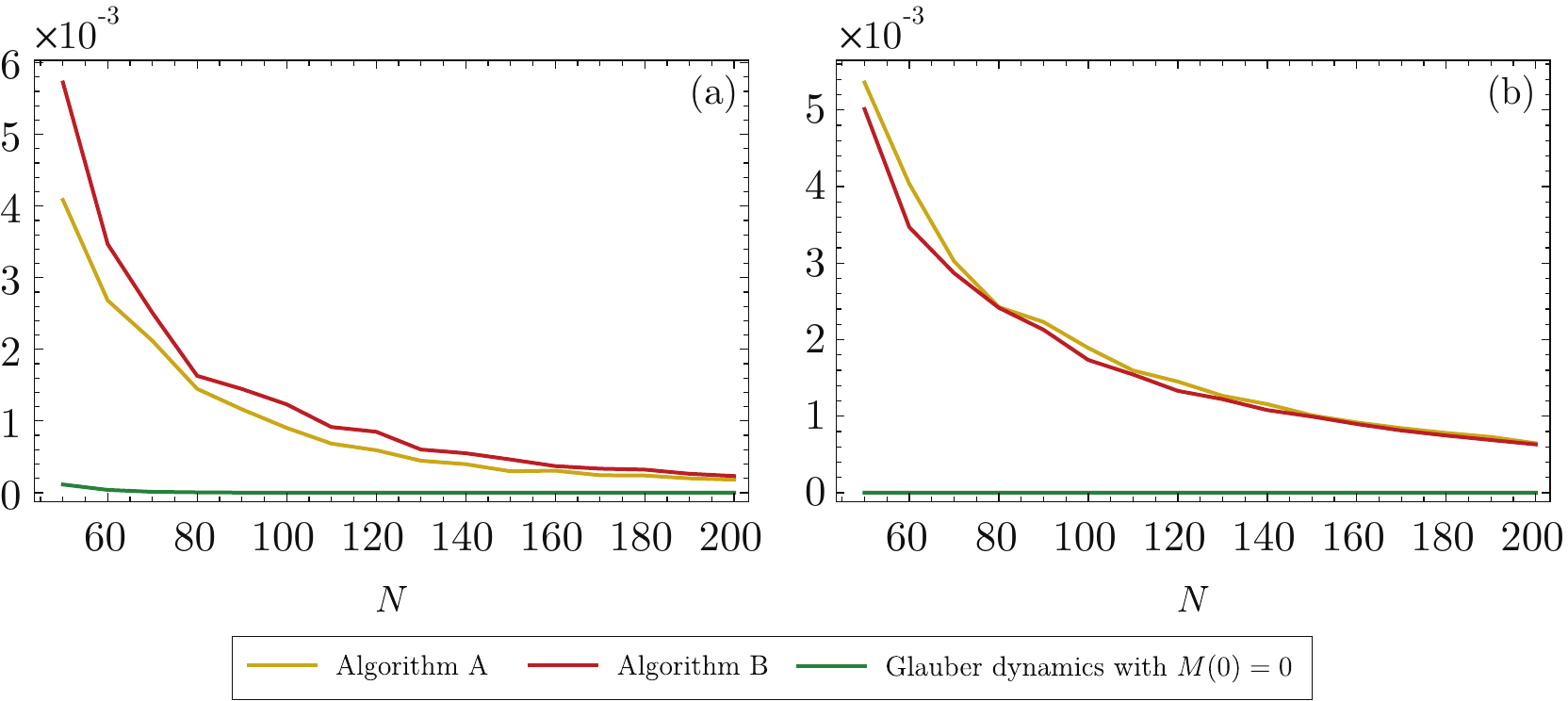} 
    \caption{The probabilities that the constrained zero-temperature search (Algorithm A), the biased  zero-temperature search (Algorithm B), and the traditional zero-temperature dynamics initialized from $M(0)=0$ absorbed into local minima under (a) $\mbox{Ber}(0.3)$ coupling distribution and (b) half-normal coupling distribution.}
	\label{fig:diffalg}
\end{figure}

The relative difference in this \emph{success probability} between the constrained and the biased zero-temperature searches seems to decrease in the $N\to\infty$ limit. See Figure~\ref{fig:diffalg} for a visualization of these differences.  On the other hand, the run time of the biased algorithm was larger by a constant factor of four to five. This longer run time is a byproduct of (a) the fact that the algorithm must first check to see if there are energy-lowering flips that also decrease the absolute magnetization before being allowed to make an energy-lowering flip that increases the absolute magnetization and (b) in instances where the algorithm fails to find traps, it must wait an extraneous order $N$ time to go from a site of magnetization $\Omega(\sqrt N)$ to the homogenous ground states. 

As such, we will presume that for $N$ large, the constrained and biased zero-temperature searches behave similarly and for the ease of numerical implementation, we only collected refined information e.g., statistics of traps, using the constrained zero-temperature dynamics.   

Additionally for ease of implementation, we actually used a variant of this algorithm in which time is counted in sweeps corresponding to $N$ single-site update attempts; the order of updates within each sweep was then chosen as a uniformly random permutation. 
We do not believe this is a crucial part of the success of the search algorithm and therefore described the dynamics as selecting one site uniformly at random for each update.  

The results presented in Section~\ref{sec:results} are for the constrained zero-temperature Glauber dynamics implemented for 100 independent realizations of the couplings $(J_{ij})_{i\neq j}\sim \nu$ and then for each such coupling realization, 
\begin{itemize}
    \item $10^5$ independent initializations and dynamical runs when $\nu$ is $\mbox{Ber}(.3)$,
    \item $10^6$ independent initializations and dynamical runs when $\nu$ is $\mbox{Ber}(.5)$ or half-normal.
\end{itemize}
We did this for values of $N$ ranging from 20 to 200, and for choices of boundary constraint $M_b = 2,4,6$. 

Finally, in Section~\ref{subsec:Cauchy}, where we compare the results to those of heavy-tailed distributions, we implemented both the traditional zero-temperature Glauber dynamics, and the constrained zero-temperature search; again these were implemented for 100 independent realizations of the couplings and $10^5$ independent initializations and dynamical runs for each such coupling realization.

\section{Results}\label{sec:results}

In this section we present and discuss numerical results pertaining to the existence and properties of local minima. The procedures and algorithms used were described in the preceding section. We separately describe two features of the models considered: in Section~\ref{subsec:finding}, we discuss the existence and algorithmic difficulty of finding local minima, and in Section~\ref{subsec:properties}, we discuss the statistics of those detected minima. We conclude in Section~\ref{subsec:Cauchy} with a comparison with the case where the edge weights are heavy-tailed.

\subsection{Finding local minima in light-tailed models}
\label{subsec:finding}
We begin by considering the central question behind this paper, namely the existence of (non-uniform) local minima in the energy landscape $\mathcal H(S)$. 
As discussed, we examine both the randomly-diluted Curie--Weiss model and the random Curie--Weiss ferromagnet with half-normal coupling distribution. We hereafter refer to the random Curie--Weiss ferromagnet with both the Bernoulli and half-normal coupling distributions as \emph{light-tailed models}.

\begin{figure}[t]
    \includegraphics[width=.8\linewidth]{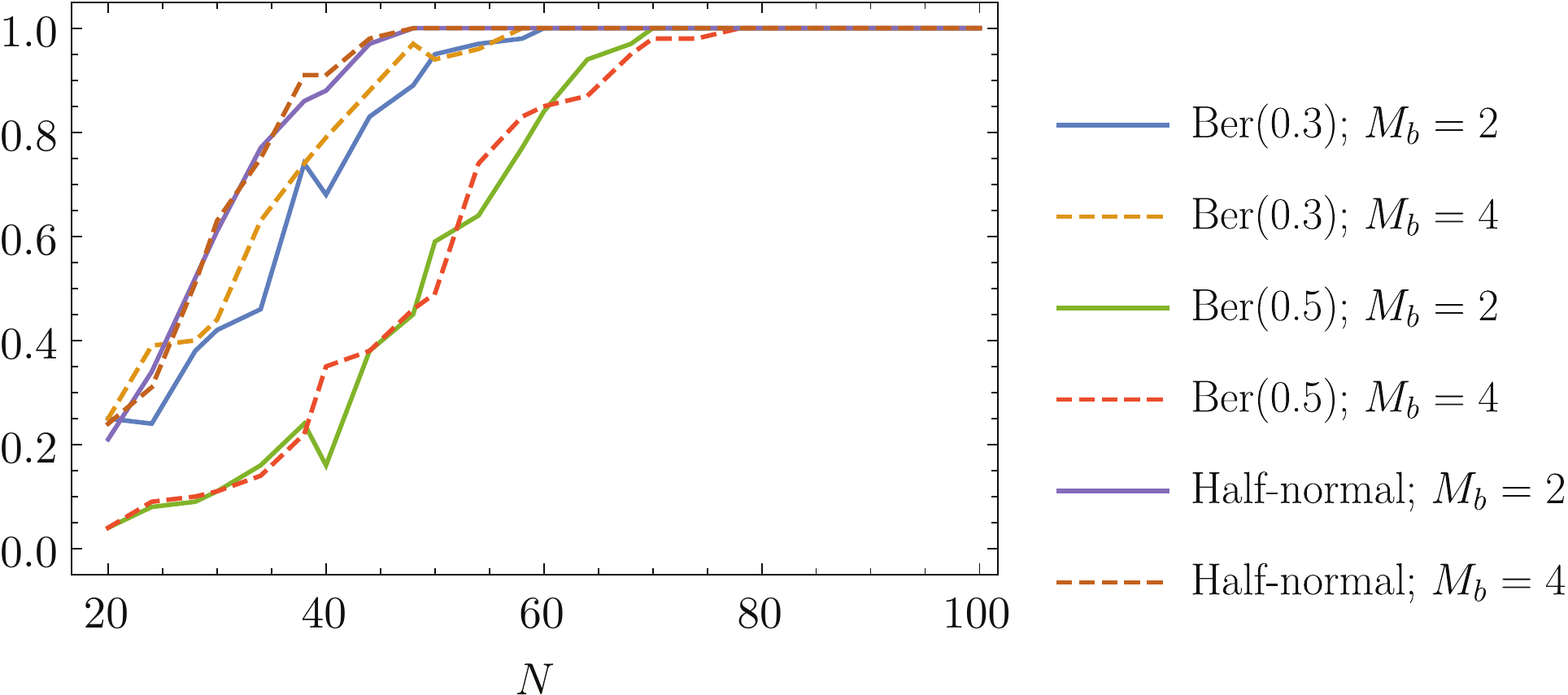} 
    \caption{Fraction of coupling realizations in which a trap was observed in at least one dynamical run of constrained zero-temperature search, for various choices of $\nu$ and $M_b$.}
	\label{fig:fracmin}
\end{figure}

Fig.~\ref{fig:fracmin} displays the fraction of coupling realizations in which at least one trap was observed, for various light-tailed models and with different magnetization boundaries. By ``fraction of graph/coupling realizations'' we mean simply the fraction of the one hundred independent realizations of the coupling collection $(J_{ij})\sim \nu$ \footnote{For smaller $N$ we considered only realizations in which the graph comprised a single connected component; otherwise the existence of a nonuniform trap would be guaranteed, but for trivial reasons.}. Note, Fig.~\ref{fig:fracmin} does not specify the fraction of dynamical runs which ended in a trap for each realization; if at least one run amongst the $10^5$ or $10^6$ absorbed into a trapping state, that graph realization possessed a trap.

Fig.~\ref{fig:fracmin} demonstrates that all the light-tailed models we investigated possess traps with high probability as $N$ gets large --- in fact, already when $N=80$, traps were detected in every one of the 100 random graph realizations. We don't see much difference between a cutoff at $M_b=2$ and $M_b=4$ in any of the cases considered, indicating that whenever a graph realization has traps, it will typically have traps of magnetization $\{0,\pm 2\}$ (in addition to possibly others of larger magnetization): indeed since the ferromagnetic drift is minimized at $M=0$, it is natural to expect that the number of local minima of a constrained magnetization is maximal at $M=0$. We explore this further in Sec.~\ref{subsec:properties}. On the other hand, it is clearly more difficult to find traps for Bernoulli distributions with $p=0.5$ than for the other two light-tailed models considered. This is not surprising given that the $p=1$ case (the ordinary Curie-Weiss ferromagnet) possesses no traps, and the strength of the ferromagnetic drift increases with $p$ (see Eq.~\eqref{eq:Hamiltonian-splitting}).

\begin{figure*}[t]

    \centering
    \includegraphics[scale=.95]{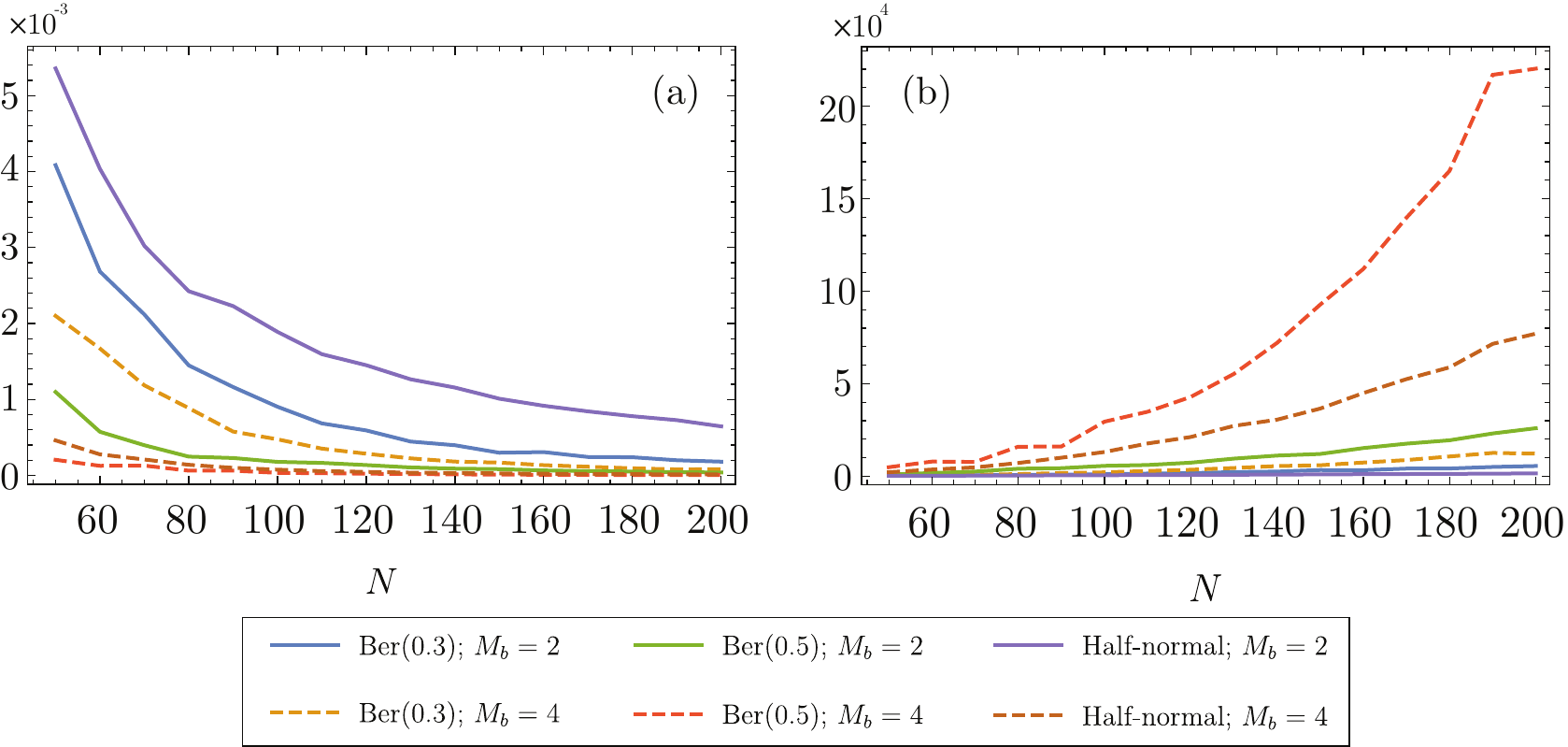}

    \caption{(a) The averaged (over coupling realizations) probability that a given dynamical run absorbs into a trap, for different models and magnetization cutoffs. (b) The inverse of the same quantity.}
    	\label{fig:succprob}
\end{figure*}

Fig.~\ref{fig:fracmin} gives no information on how likely a given algorithmic run is to find a trap. This is addressed in Fig.~\ref{fig:succprob}, which examines the probability of success of a given run finding a trap.  This probability is estimated by counting the number of runs absorbed in a trap among $10^5$ total runs for each realization for Ber(.3) and $10^6$ runs per realization for Ber(.5) and half-normal; the result is then averaged over 100 realizations per model.  

The figure makes clear the difficulty of finding traps as $N$ grows, which perhaps is not surprising given the exponential growth of the configuration space; indeed a simple calculation suggests that the expected number of traps is an exponentially decaying fraction of $\{\pm 1\}^N$ (though possibly still growing exponentially in $N$). In part (b) of the figure, we show the behavior of the inverse probability as $N$ gets large. Within numerical error it appears to be growing polynomially  with~$N$ (see discussion in Sec~\ref{sec:discussion}).

\begin{figure}[t]
    \centering
    \includegraphics[scale=0.6]{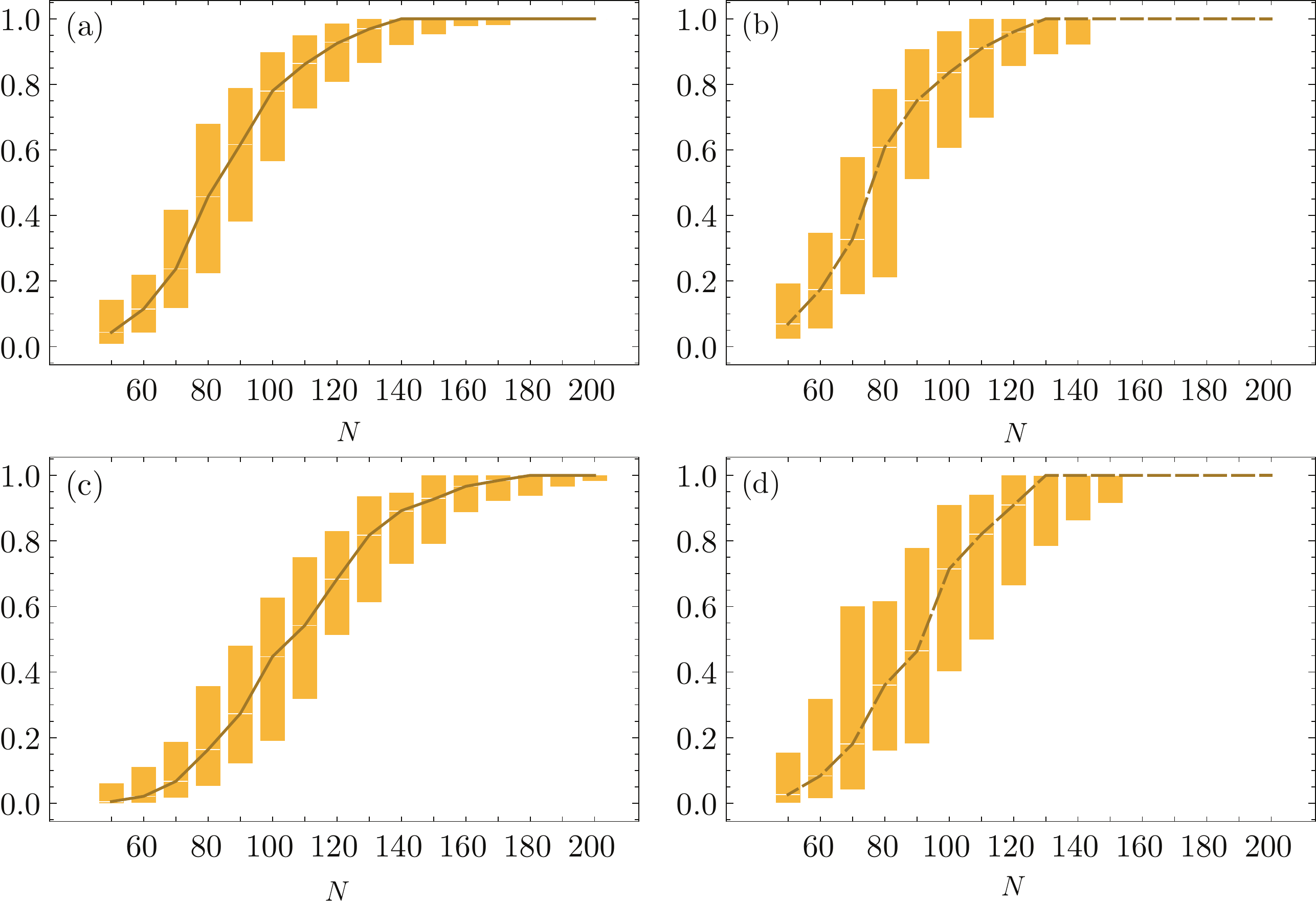}
    \caption{The ratio of the number of distinct traps observed to the number of runs absorbed in a trap, for (a) $\mbox{Ber}(0.3)$ disorder and $M_b=2$. (b) $\mbox{Ber}(0.3)$ disorder and $M_b=4$. (c) Ber$(0.5)$ disorder and $M_b=2$. (d) Ber$(0.5)$ disorder and $M_b=4$. The solid/dashed lines indicate the median; the error bars contains $80\%$ of the data. }
    \label{fig:distinct}
\end{figure}

As indicated above, the natural next question is whether one can obtain some estimate on the number of distinct traps. While precise estimates on this would require an exponential number of queries, we can use the number of  repeated dynamical observations of the same trap as a proxy. Namely, if the number of traps were rapidly diverging as $N\to\infty$, and they were roughly equally likely to absorb a given dynamical run that ends in a trap, independent dynamical runs should almost always end in distinct traps. Indeed this is the scenario we find in Fig.~\ref{fig:distinct}, which shows that for $N\gtrapprox 150$ essentially all traps found by the dynamical searches are distinct. This indicates that the number of traps (even restricted to those of magnetization $\{0,\pm 2\}$ or $\{0,\pm 2, \pm 4\}$) is diverging as $N\to\infty$. 

\begin{figure}
    \centering
    \includegraphics[scale=0.84]{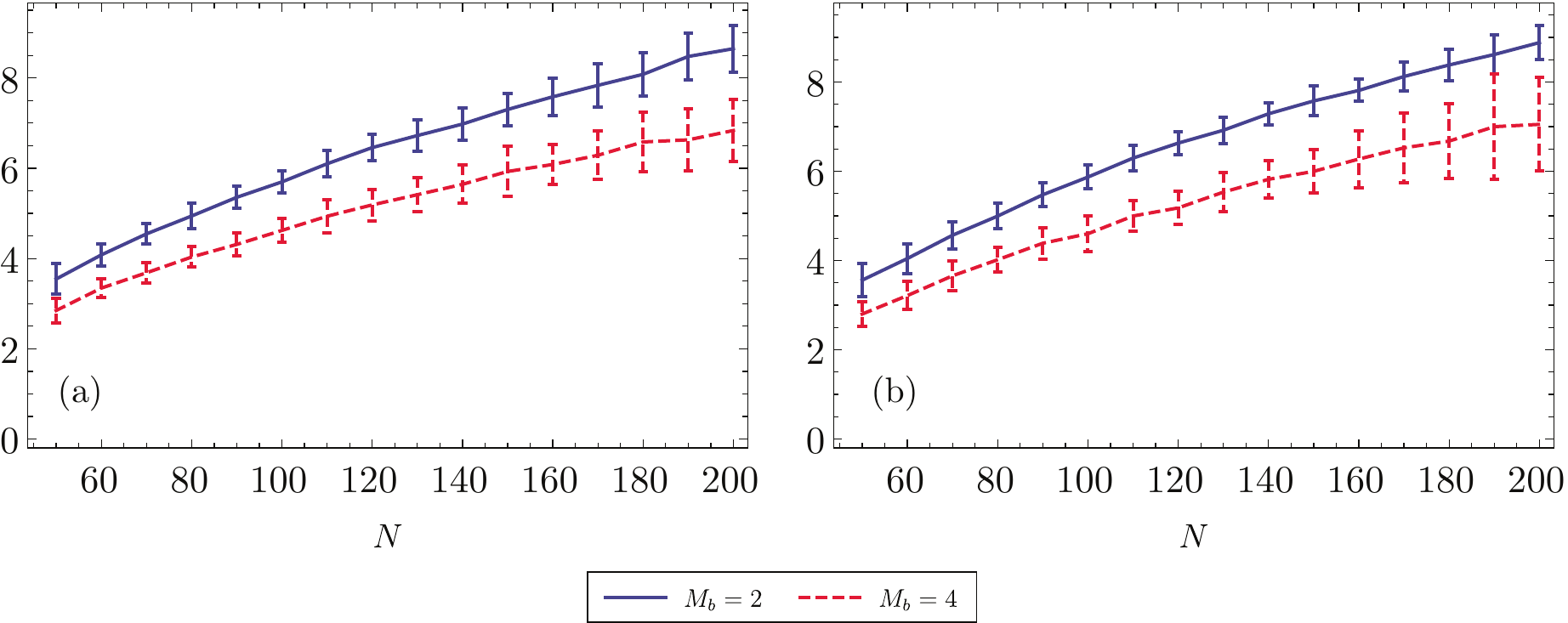}
    \caption{Average number of time sweeps (as described in Sec.~\ref{sec:level2}) to absorption, when the dynamics ends in a trap under (a) Ber$(0.3)$ disorder (b) Ber$(0.5)$ disorder. Error bars denote one standard deviation.}
    \label{fig:time}
\end{figure}

Given that every update in the algorithm decreases the energy, one would expect that in a run that does find a trap, the time to get there is relatively short. Fig.~\ref{fig:time} (a)--(b) provide information on this for the two Bernoulli models examined here for different magnetization cutoffs. While some variation is seen among the different models and cutoffs, it appears to be relatively small. 

\begin{figure}
    \centering
    \includegraphics[scale=0.84]{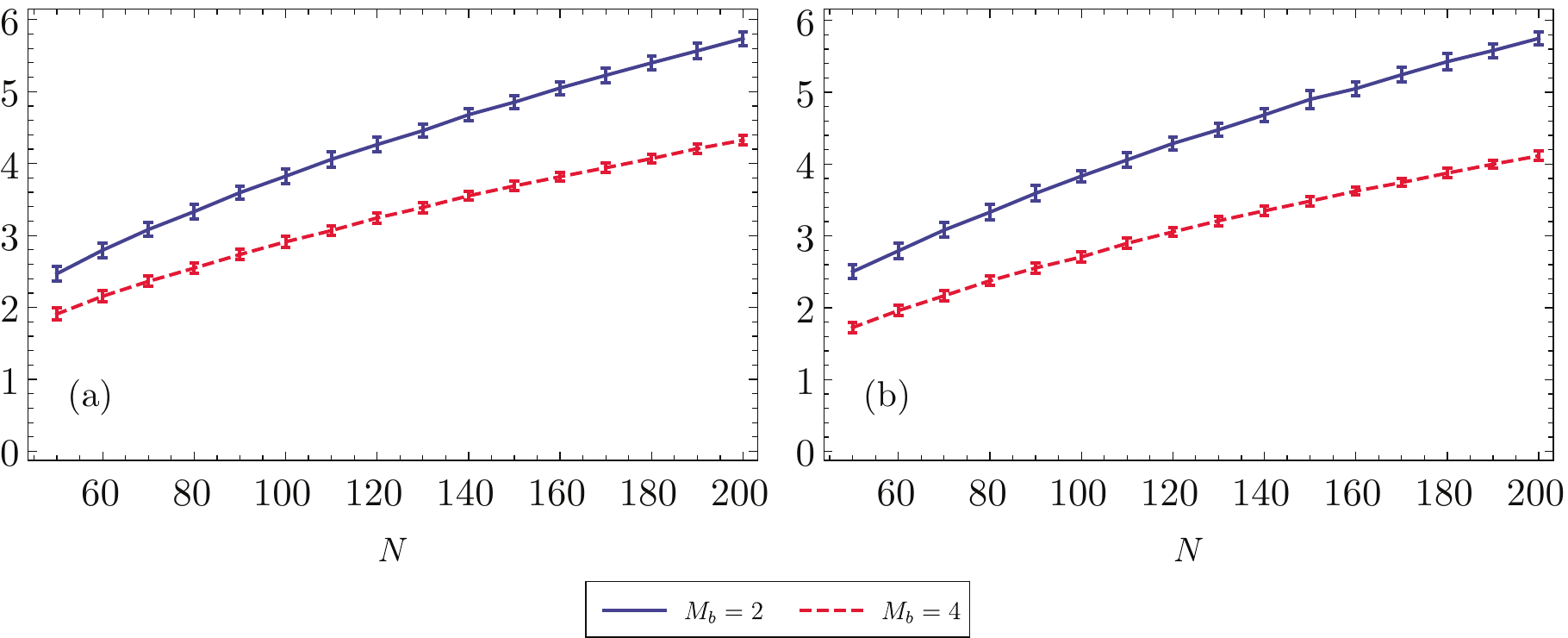}
    \caption{Average number of time sweeps until absorption when the dynamics fails to find a trap under (a) Ber$(0.3)$ disorder (b) Ber$(0.5)$ disorder.}
    \label{fig:boundary}
\end{figure}

As seen earlier, most runs for Bernoulli and half-normal models are absorbed by the cutoff boundary rather than a trap. Thus for purposes of comparison we show in Fig.~\ref{fig:boundary} the average times to be absorbed at the boundary for runs that don't find a trap. Again, the differences in absorption times are noticeable but relatively small between traps and boundary points wherein the only lower-energy neighbors have larger magnetization. In particular, both timescales are increasing sublinearly with $N$ in the range studied, and it appears that both timescales (when viewed as number of sweeps) are slowly diverging; but it cannot be ruled out that the timescales converge to a finite value at longer times.

\subsection{Statistics of local minima}
\label{subsec:properties}

The results in Sec.~\ref{subsec:finding} provided numerical evidence that with high probability as $N\to\infty$, traps exist in both the Bernoulli and half-normal random ferromagnets, despite the (rigorous, for the Bernoullli case) results of~\cite{GNS18}  that standard energy-lowering 1-spin-flip dynamics at zero temperature lead, with very high probability, to one of the two uniform states. Unfortunately, due to the relatively sparsity of traps (an exponentially small fraction of the space) and the difficulty of finding them, we are not able to obtain reliable results on how fast the number of distinct traps grows with $N$. 

Continuing the investigation of the landscape structure of these models---which we have found to be complex (i.e., consisting of many local minima) at near-zero magnetization---we study properties of these traps. Using the traps into which the dynamics absorbs as a proxy for typical traps (our initialization is independent of the landscape) we analyze the distributions of the energy and magnetization of traps, and in the Bernoulli case we also study the distribution of the volumes (i.e., number of spin configurations) of traps.

\begin{figure*}[t]
    \centering
    \includegraphics[scale=0.61]{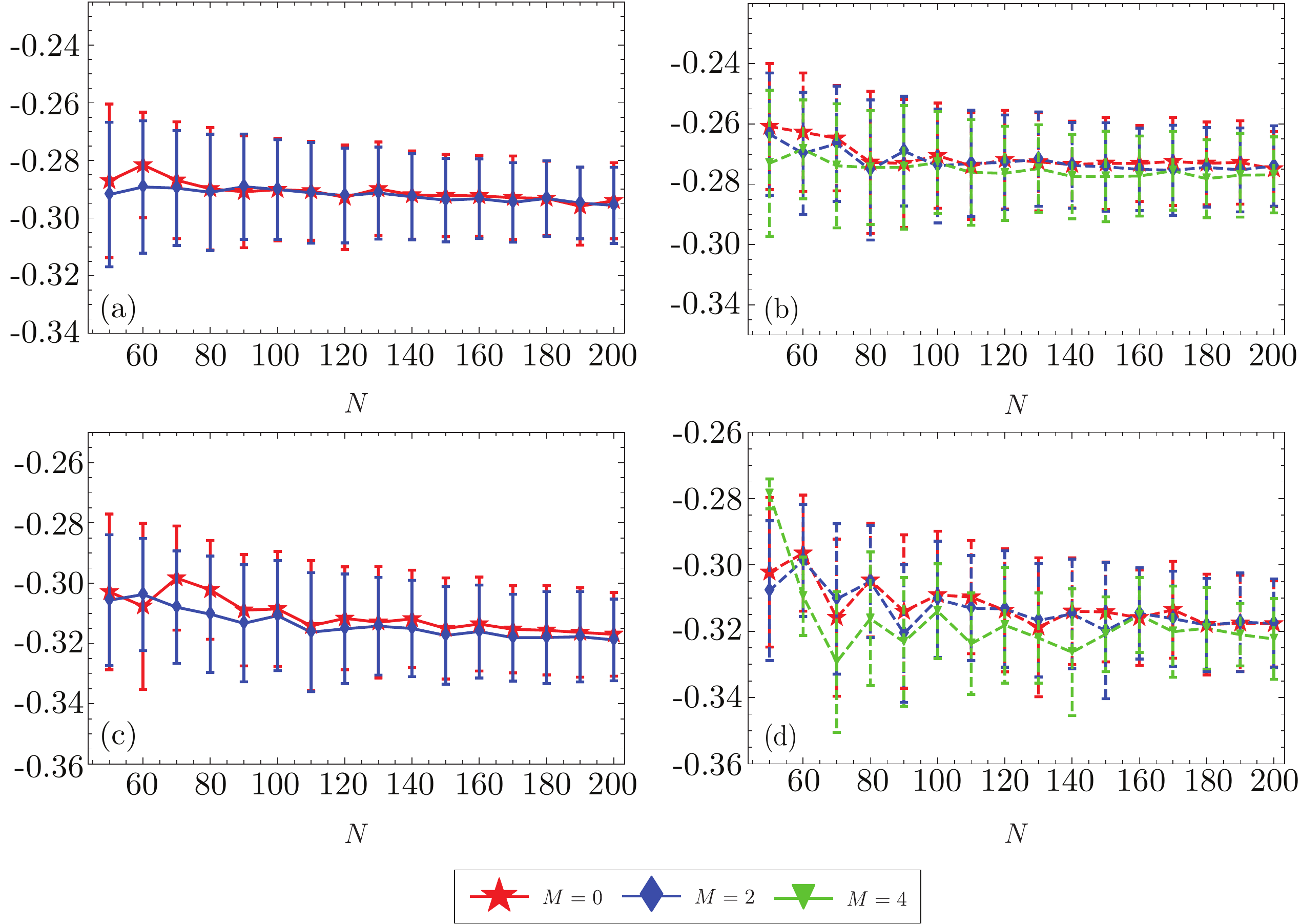}
    \caption{Average rescaled (divided by $N^{3/2}$) energy of  absorbing traps having magnetization $0,2,4$ under  
    (a) Ber$(0.3)$ disorder and $M_b =  2$ (b) Ber$(0.3)$ disorder and $M_b=4$ (c) Ber$(0.5)$ disorder and $M_b =2$ (d) Ber$(0.5)$ disorder and $M_b =4$. 
    }
    	\label{fig:energy}
\end{figure*}

Fig.~\ref{fig:energy} shows the rescaled energy of trap configurations vs.~$N$. (Note that for these models at magnetizations that are $O(1)$, the natural rescaling of the energy is $N^{3/2}$; for a uniform Curie-Weiss ferromagnet, as in for random ferromagnets at $O(N)$ magnetizations, the total energy would be divided by $N^2$.) 
We see from the figure that the rescaled energy is remarkably constant as $N$ increases, indicating that the large-$N$ limit for this quantity has already been reached. In addition, one can see that the result is insensitive to the cutoff magnetization used in the algorithm. It therefore appears that the typical trap energy divided by $N^{3/2}$ converges to some limiting constant as $N\to\infty$. 

\begin{figure*}[t]
    \centering
    \includegraphics[scale=0.63]{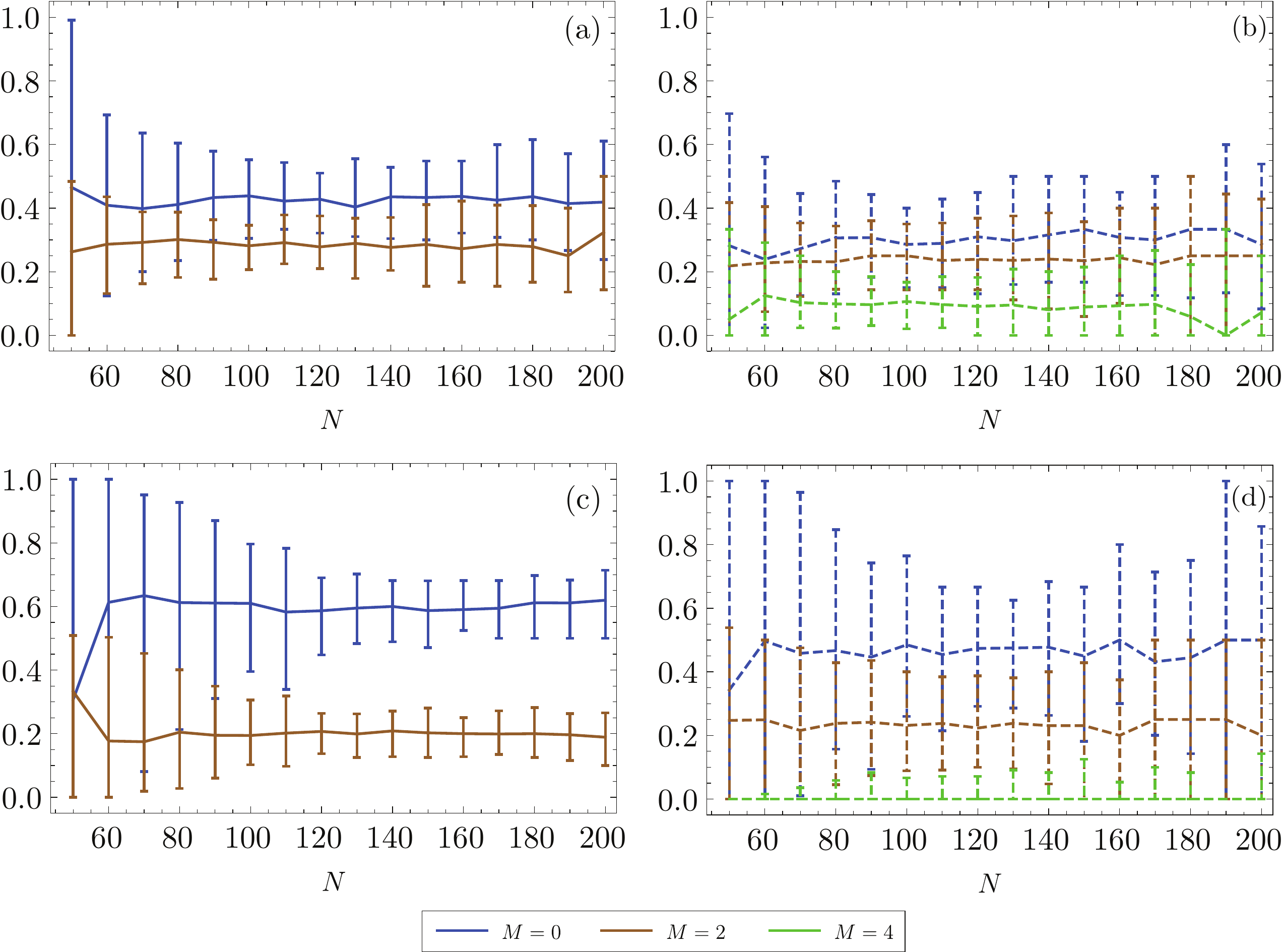}
    \caption{Fraction of dynamical runs ending up in traps of magnetization $M=0,2,4$ under (a) Ber$(0.3)$ disorder and  $M_b=2$ (b) Ber$(0.3)$ disorder and $M_b=4$ (c) Ber$(0.5)$ disorder and  $M_b=2$ (d)  Ber$(0.5)$ disorder and $M_b=4$. Error bars contain $80\%$ of the data.}
    	\label{fig:mag}
\end{figure*}

We next turn to the magnetization of the traps detected. For all runs that are absorbed in a trap, Fig.~\ref{fig:mag} shows the probability of landing in a trap with magnetization $M$.  The figure demonstrates that traps can have varying magnetizations: $0,\pm 2,\pm 4...$. (In Fig.~\ref{fig:mag}, only positive magnetizations are shown. By spin-flip symmetry the distributions of magnetizations of local minima are symmetric and this was confirmed to within numerical error.)

The figure shows that the frequency of traps of each magnetization found by the algorithm depend on the cutoff magnetization used, which should not be surprising: e.g., traps with magnetization $\pm 4$ cannot be found by an algorithm with an absorbing boundary of magnetization $\pm 2$, and so on. However, for a fixed boundary $M_b$, the numerics appear consistent with the hypothesis that traps of magnetization $0$ are the most prevalent (there is both the most configuration-space volume and the minimal ferromagnetic drift at that magnetization), and that this prevalence decreases to zero as the magnetization increases on $O(1)$ scales. 

Contrast this with the results (cf.~Fig.~\ref{fig:energy}) for trap energies, which are (to within numerical error) independent of trap magnetization or cutoff used. This is analogous to a result proved in~\cite{NS99c} for finite-dimensional random ferromagnets: all 1-spin flip metastable states in fixed dimension have the same energy per spin w.p.~1. The same would be true for magnetization per spin (recall that Fig.~\ref{fig:mag} shows the total magnetization, not the magnetization per spin). It remains a conjecture as to whether these results carry over to their natural finite-size analogues in infinite-range models, but the numerics shown here indicate that they do. Namely, they indicate that the laws of the rescaled energy as well as the magnetization per spin of a local minimum chosen uniformly at random converge to Dirac masses.

\begin{figure}[t]
    \centering
    \includegraphics[scale=0.6]{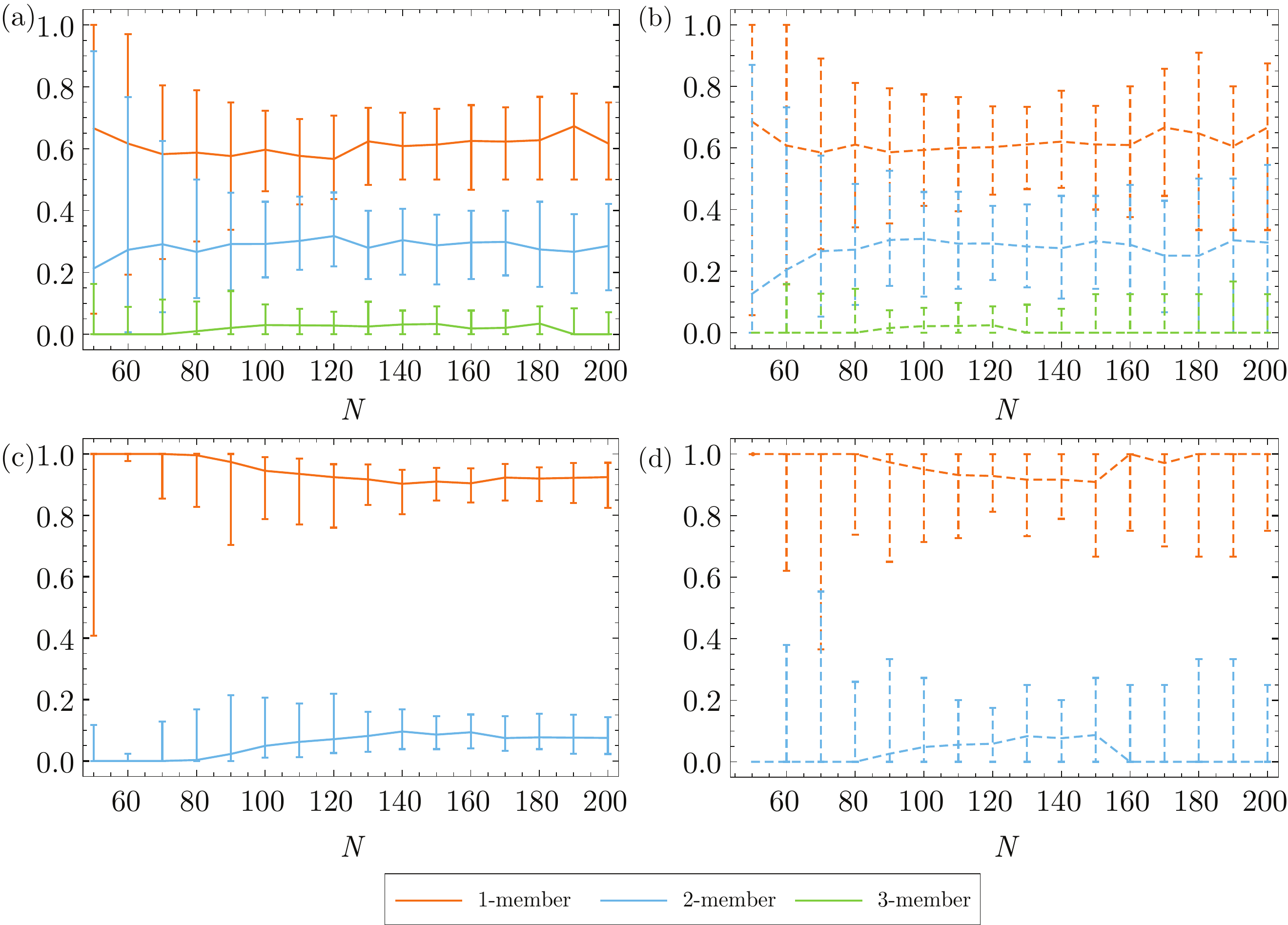}
    \caption{Fraction of dynamical runs absorbing into traps having $j$-members for $j=1,2,3$ for (a) Ber$(0.3)$ disorder and $M_b=2$ (b) Ber$(0.3)$ disorder and $M_b=4$ (c) Ber$(0.5)$ disorder and $M_b=2$ (d) Ber$(0.5)$ disorder and $M_b=4$. The error bars contain 80$\%$ of the data.}
    \label{fig:j}
\end{figure}

Fig.~\ref{fig:j} provides information on the number of distinct spin configurations constituting the individual traps --- i.e., the number of configurations such that one can move between them by paths with only zero-energy flips. In models whose disorder distribution is continuous, e.g., the half-normal distribution, the probability of neighboring configurations having the same energy is zero, and thus all traps consist of exactly a single spin configuration. This is not the case for models with $\mbox{Ber}(p)$ disorder, however, where the disorder distribution is atomic and adjacent spin configurations can have equal energy.  In Fig.~\ref{fig:j} we find that some traps do indeed consist of connected components of such configurations, with traps of up to three equal-energy spin configurations detected; unsurprisingly, the frequency of traps with increasing numbers of spin configurations falls fairly rapidly. It seems likely that ``$j$-member traps'' with $j>3$ exist, but they become increasingly difficult to find using our algorithm.

\subsection{Heavy-tailed distributions}
\label{subsec:Cauchy}

\begin{figure*}[t]
    \centering
    \includegraphics[scale=0.65]{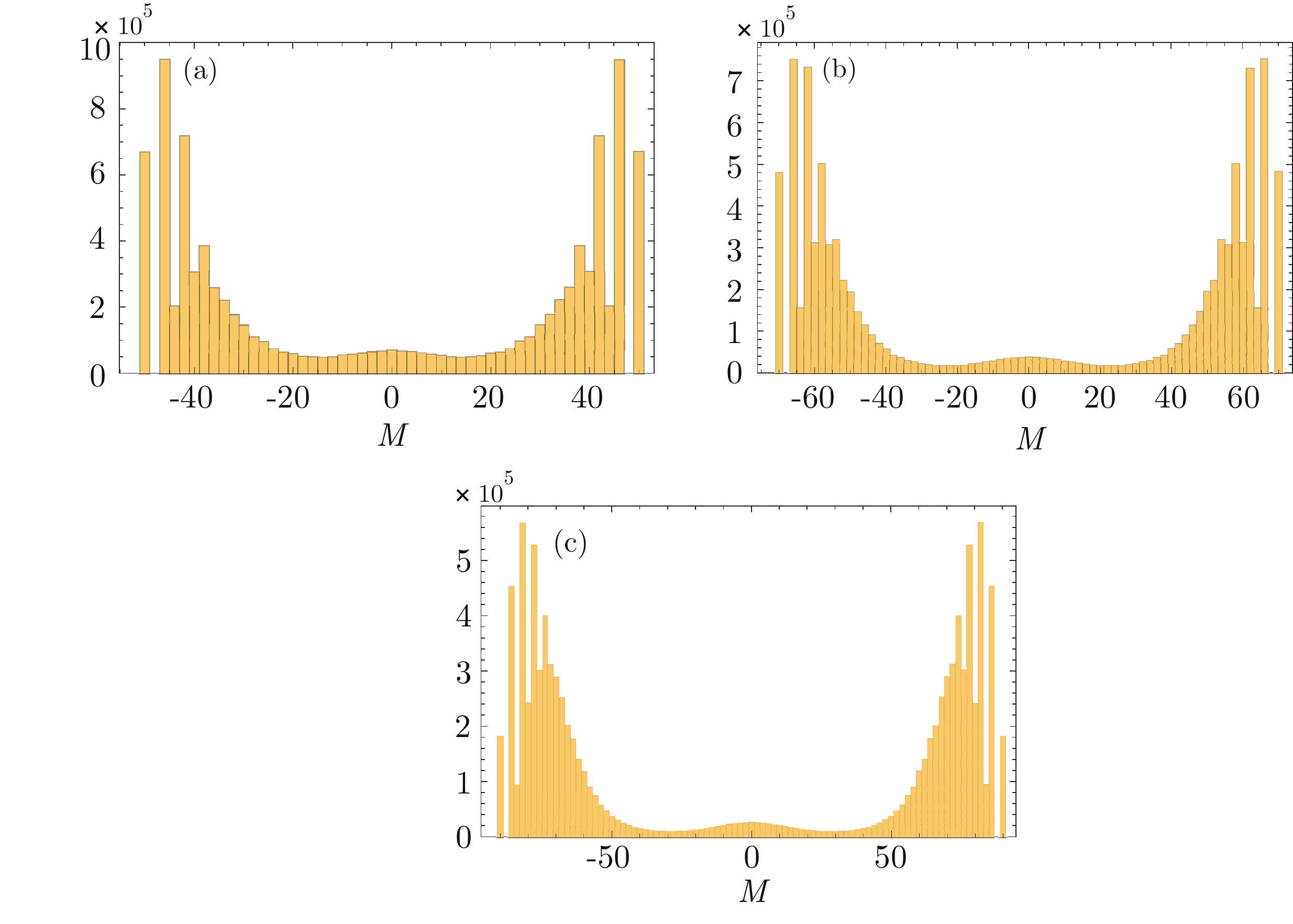}
    \caption{Final magnetization distributions for the half-Cauchy model with (a) $N=50$, (b) $N=70$, and (c) $N=90$ using the traditional zero-temperature Glauber dynamics.}
    	\label{fig:hist1}  
\end{figure*}

For the sake of comparison, we also considered the case of a half-Cauchy distribution~(\ref{eq:Cauchy}) (having an upper tail that decays as $1/(\pi[1+x^2])$). 
The analysis in~\cite{GNS18} proved (in the case of power-law distributions with upper tail that decays slower than $1/x^2$) that heavy-tailed models have exponentially (in $N$) many 1-spin-flip stable traps, and with probability bounded away from zero as $N\to\infty$ the ordinary zero-temperature Glauber dynamics absorbs into a metastable trap.

\begin{figure*}[t]
    \includegraphics[scale=0.63]{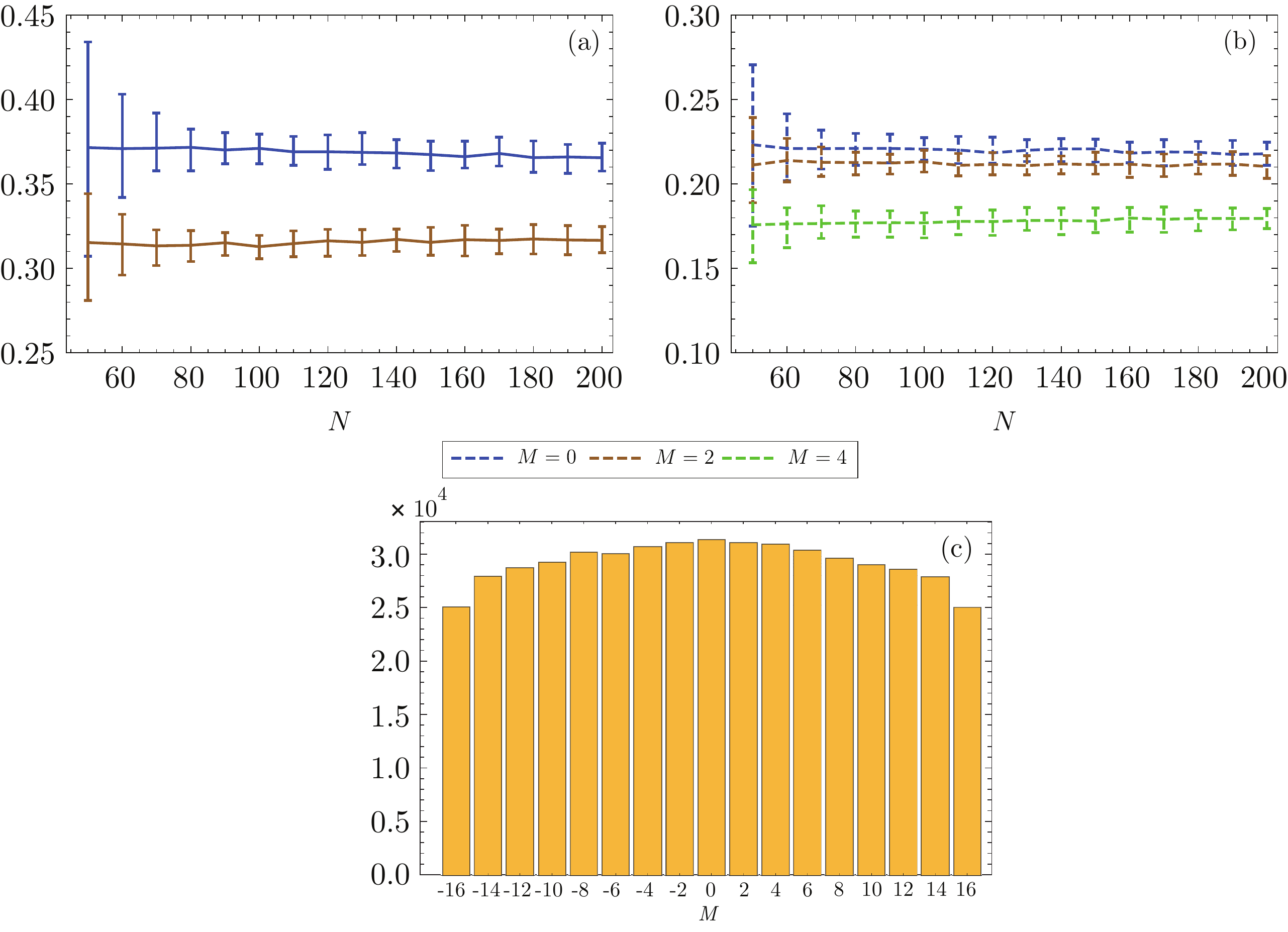}
     \caption{Distribution of final magnetizations for the half-Cauchy model with $N=100$ under constrained zero-temperature dynamics at (a) $M_b=2$ (b) $M_b=4$ (c) $M_b=16$.
    The error bars contain 80\% of the data around the median.}
    	\label{fig:hist2}
\end{figure*} 

Fig.~\ref{fig:hist1} shows histograms of the magnetization distribution of observed traps in the half-Cauchy model for three different values of~$N$. We note that as $N$ increases the bulk of the weight of the distribution shifts to higher absolute magnetizations of order $N$. It is not the case, however, that as $N\to\infty$, the distribution evolves to a pair of $\delta$-functions at $\pm 1$ (which would be the case for light-tailed distributions); numerical results from~\cite{GNSW} (see in particular Fig.~5 of that paper) demonstrate that as $N\to\infty$ the (magnitude of the) averaged final magnetization per site levels off at values strictly below one.

In order to more closely study the magnetization distribution at small~$M$, we introduced cutoffs. The results are shown in Fig.~\ref{fig:hist2}. The distribution as seen in part~(c) of the figure is peaked at zero but slowly varying in this range. The fraction of dynamical runs of the constrained zero-temperature chain absorbing into a trap was higher than the corresponding fraction in light-tailed models by a factor of ten. Thus it seems that in these heavy-tailed models, traps of near-zero magnetization are much more plentiful and/or more accessible than in the corresponding light-tailed models. The relevant mean-field effects in the two are the same, but the absence of concentration in heavy-tailed models induces a much rougher landscape that does not get smoothed out by the ferromagnetic drift.  

\begin{figure}[t]
    \includegraphics[scale=0.6]{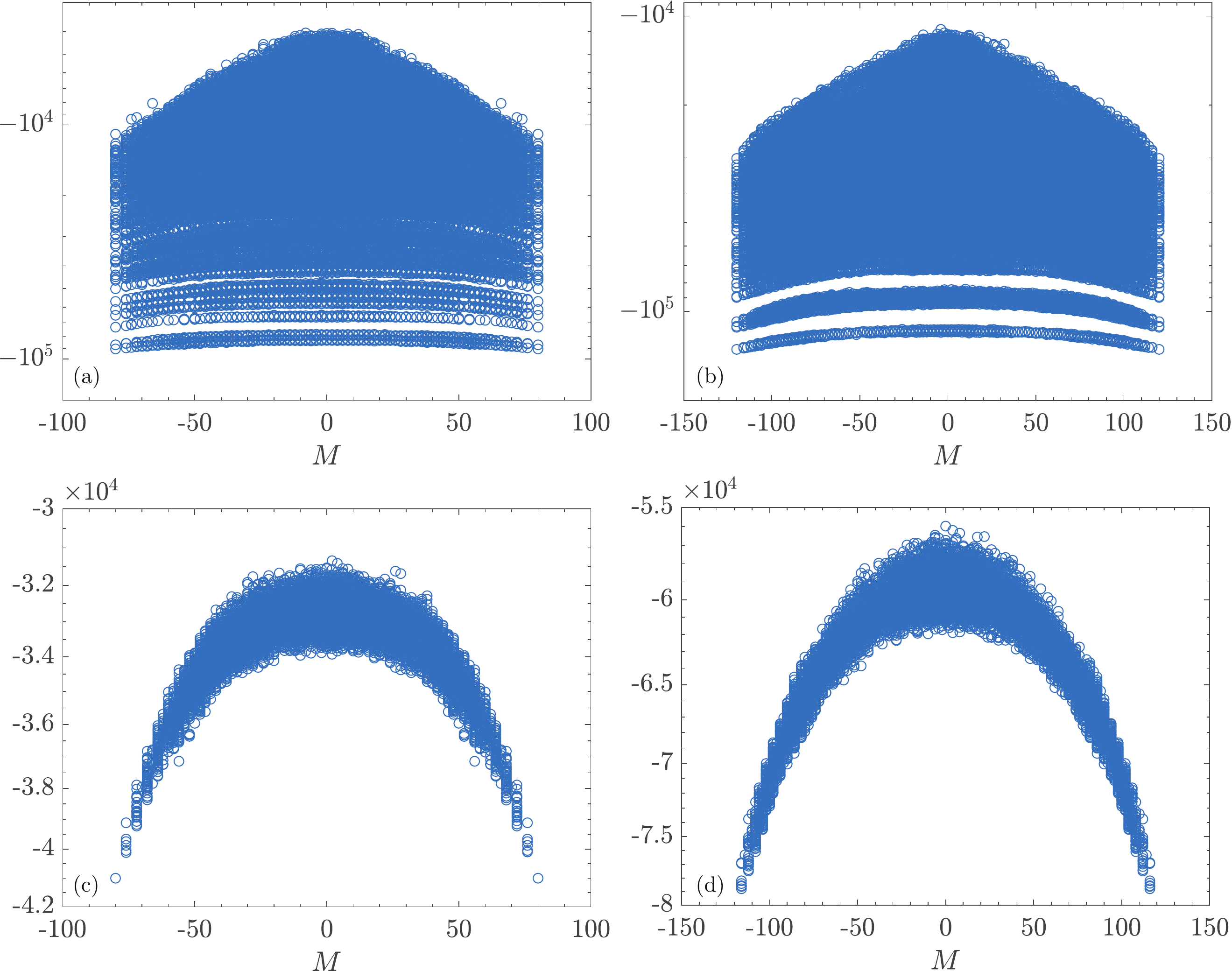}
    \caption{Scatterplots of energy versus magnetization of observed traps for the half-Cauchy model, at $N=80$ and $N=120$. (a) 100 coupling realizations; $N = 80$.; (b) 100 coupling realizations; $N = 120$.; (c) One coupling realization; $N=80$.; (d) One coupling realization; $N = 120$.}
    \label{fig:scatterplot}
\end{figure}

Finally, we examined the relationship between energy and magnetization of observed local minima in the landscape of heavy-tailed random ferromagnets. In Figure~\ref{fig:scatterplot}, we depict scatterplots of the energy and magnetization of distinct local minima trapping the standard zero-temperature dynamics. Due to the non-concentration of the heavy-tailed disorder, different coupling realizations lead to different bands of energy levels for their corresponding set of traps. When focusing on a particular coupling realization, though, it becomes clear that the order of magnitude of the deeper energies induced by the heavy-tailed disorder appears to be on the same scale as that induced by the ferromagnetic drift. Again, this is in stark contrast with the light-tailed situation, wherein the energy of the global  minimum is on a different scale than that of the deepest local minima with small magnetization.

\section{\label{sec:discussion}Discussion}
In this section, we summarize the results of Section~\ref{sec:results}, then discuss them from various perspectives arising from the existing  literature and propose a series of open questions in the understanding of the landscapes of random mean-field ferromagnets. As mentioned in the introduction, there are several different lines of research to which the models we study may be of interest, most pertinently: 
\begin{enumerate}
\item Complexity of spin glasses and random landscapes
\item Constrained optimization problems in theoretical computer science
\end{enumerate} 

Before discussing the relationship between our findings and these two research perspectives, let us summarize our numerical findings.

\subsection{Summary of results}
Our aim is to characterize the energy landscape of disordered mean-field ferromagnets, including randomly diluted (``Bernoulli") models; models with light-tailed distributions (e.g., half-Gaussian); and models with fat-tailed distributions (e.g., half-Cauchy). The general behavior appears to separate into two general classes of behavior, with Bernoulli and light-tailed models showing one type of behavior, and fat-tailed models the other. (For this reason, ``light-tailed" in the following discussion will include Bernoulli.)

The behavior of fat-tailed random Curie-Weiss models is qualitatively similar to what one encounters in the Sherrington-Kirkpatrick~\cite{SK75} spin glass model and in short-range random ferromagnets and spin glasses in finite dimensions: traps are plentiful and standard zero-temperature Glauber dynamics lands in a trap with probability approaching one as $N\to\infty$. In Section IV.C we studied these models and characterized the magnetization distribution of their traps. This additionally served as a test of the algorithms we introduced to study the light-tailed models, to which we now turn. 

The energy landscape of light-tailed random Curie-Weiss models is the focus of this paper. As read off from~\eqref{eq:Hamiltonian-splitting}, the disorder in the landscapes competes with a quadratic Curie--Weiss potential; the latter becomes dominant at diverging magnetizations.  
As a consequence, it was proven in~\cite{GNS18} that the usual method of finding traps, namely observing the dynamical evolution according to zero-temperature Glauber dynamics starting from a randomly chosen spin configuration, is unlikely to find traps: indeed from Figure~\ref{fig:diffalg} it appears this probability decays exponentially in $N$ as $N\to\infty$. The theoretical results of~\cite{GNS18} and numerical runs of traditional zero-temperature dynamics raised the question of whether these landscapes have \emph{any} local minima.

In order to resolve this question and obtain some preliminary understanding of traps in these models, we introduced two new algorithms designed specifically to search for traps in such landscapes. The first is a constrained zero-temperature dynamics algorithm, in which (roughly speaking) a reflecting hard boundary is imposed at an order one absolute magnetization; the idea is that once the dynamics evolves past that point there will almost certainly follow a quick downhill run to the corresponding uniform all-plus or all-minus state. The second algorithm introduces a biased zero-temperature dynamics. Here there are no ``hard" walls but rather an imposed drift pushing the system toward smaller absolute magnetization whenever possible.

The two algorithms gave essentially the same results. They clearly demonstrated that light-tailed models do indeed possess (non-uniform) local minima, and the number of such local minima diverges with $N$ (though apparently more slowly than in heavy-tailed ferromagnets and spin glasses). Moreover, we were able to study the statistics of these local minima; i.e., distributions of magnetizations, normalized energies, and in the Bernoulli case (where energy ties can occur) the distribution of the number of equal-energy spin configurations forming the traps. Detailed results are given in Sect.~\ref{subsec:finding} and \ref{subsec:properties}.

\subsection{Complexity of disordered landscapes}
One active research direction into which our results fall is the complexity of random landscapes in high dimensions. A common theme in the study of mean-field spin glasses (Hamiltonians of the form~\eqref{eq:energy} with $\nu$ symmetric rather than non-negative) is the number and structure of their critical points. When the state space of mean-field spin glass landscapes is relaxed from the hypercube to the $(N-1)$-sphere $\mathbb S^{N-1}(\sqrt N)\supset \{\pm 1\}^N$, a canonical way of defining complexity is via the number of critical points and/or local minima of $\mathcal H$. In that setting, it is a hallmark of random landscapes that the number of local minima (and more generally the number of critical points of every index) diverges exponentially in $N$~\cite{ABA13,ABC13,Sub15}. 
On the hypercube, the analysis is different, as there is no access to powerful tools like the Kac--Rice formula: see~\cite{ADLO} for a computation of the expected number of 1-spin flip stable states of the Sherrington--Kirkpatrick model.

On the sphere, spin-glass models with a quadratic spike $-\lambda (M(S))^2$ (which as noted earlier in~\eqref{eq:Hamiltonian-splitting} is close to our setup) have been studied at the annealed level. With $p$-spin interactions and Gaussian disorder, the expected number of local minima has been analyzed as a function of the magnetization and found to undergo a transition as the strength of the quadratic spike varies. In particular, as the spike strength $\lambda$ increases, the landscape trivializes so that there are no critical points besides the ground states above a threshold magnetization $M_{triv}(\lambda)$; however, it is expected (and confirmed at the annealed level) that there are  many critical points with $|M(S)|<M_{triv}(\lambda)$. As in our setting, there is a close relation between this complexity picture and the trapping of gradient-based algorithms as well as their success in signal recovery for spiked tensors~\cite{MR14,BMMN17,BAGJ18,BBCR18}.  

When restricting to the hypercube and the case of two-spin interactions, we arrive essentially at the ferromagnetic setting considered in this paper.  As we saw in Section~\ref{subsec:finding} (in particular, Figure~\ref{fig:fracmin}), the existence of local minima in the energy landscape with probability going to $1$ as $N\to\infty$, was confirmed numerically. As such, despite the triviality of the Glauber dynamics on $\mathcal H(S)$, the landscape itself does not completely smooth out, and retains some of the complexity of related disordered systems. 

In light of the context above, and the fact that numerics constrain us from obtaining a good understanding of the multiplicity of local energy minima (beyond the fact that this number diverges as indicated in Figure~\ref{fig:distinct}), we pose the following natural questions:    

\begin{question}
How fast does the number of local minima of disordered mean-field ferromagnets, $\mathcal H(S)$, with light-tailed disorder (say Bernoulli or half-normal) diverge as $N\to\infty$? Does it grow exponentially with $N$, or more slowly?
\end{question}

We were able to get some grasp on the structure of local minima in terms of typical magnetizations and energies, as observed by our algorithms. Of course, it begs the question of whether the law of minima observed by our algorithm is close to that of a local minimum chosen uniformly-at-random. Motivated by the topological transition in the complexity of spiked $p$-spin glasses described above, as well as the results in Figures~\ref{fig:energy}--\ref{fig:mag}, we pose the following questions: 

\begin{question}
For a given coupling distribution $\nu$ with mean $\lambda$, what is the smallest magnetization $M^\epsilon_{triv}(\lambda)$ such that with probability $\ge 1-\epsilon$ as $N\to\infty$, the only local minima with $|M(S)|\geq M^\epsilon_{triv}(\lambda)$ are the all-plus and all-minus states? Is it the case that $M^\epsilon_{triv}(\lambda)$ is of order one in~$N$ for any fixed~$\epsilon$? 
\end{question}

\begin{question}
Let $\mathbf P$ be the law of a local minimum $S^\star$ drawn uniformly at random from the local minima of $\mathcal H$. Does $M(S^\star)$ under $\mathbf P$ converge as $N\to\infty$ to a non-degenerate law? Does $\mathcal H(S^\star)/N^{3/2}$ under $\mathbf P$ converge to a point mass as $N\to\infty$? 
\end{question}

The tightness of the magnetization described above is conjectural but seems consistent with the stability of the magnetization of observed local minima in Figure~\ref{fig:mag}. This would suggest that the order one cutoff $M_b$ for the constrained Glauber dynamics we proposed is the correct scaling for the boundary in order to find the local minima of $\mathcal H$. 

Looking in the vector space orthogonal to the $(1,...,1)$-direction in which a quadratic potential is imposed by the ferromagnetic constraint~\eqref{eq:Hamiltonian-splitting}, one can further delve into the structure of these local minima by examining a canonical quantity known as the \emph{overlap}. 

\begin{question}
Consider two independently, uniformly chosen local minima of $\mathcal H$ under $\mathbf P$, $S^1, S^2$. Does the law of their overlap $\frac 1N \sum_i S^1_i S^2_i$ converge to some limit? Is the limiting law a single point mass at zero? During
a dynamical search $S(t)$ starting from a random initial spin configuration
$S(0)$, what can be said about the time-dependent overlap between $S(t)$ and $S(0)$? 
\end{question}

\subsection{Relation to combinatorial optimization}

Recently, there has been intensive research activity focusing on the landscape structures and algorithmic tractability of random instances of constrained optimization problems. A series of classical examples of this are extremal cut problems on random graphs, or more generally, the complete graph with random edge-weights drawn i.i.d.\ from some measure $\nu$.  

In~\cite[Section 1.1]{GNS18}, the local minima of random mean-field ferromagnets were recontexualized as \emph{local MINCUT} problems of a corresponding random graph. Namely, if we take the complete graph $K_N$ with edge weights $J_{ij}$, a \emph{local MINCUT} is a partition $(A,A^c)$ of the vertices $\{1,\ldots,N\}$ such that its cut-width $CUT(A) = \sum_{1\leq i<j\leq N} J_{ij}$ is at most that of any neighboring partition $(B,B^c)$ having Hamming distance $d_H(A,B)= 1$.

Extremal cuts (e.g., MAXCUT, MIN bisection) are fundamental objects in the theory of randomized optimization and constraint satisfaction that are widely believed to be algorithmically intractable with high probability. Important progress towards understanding the landscape complexity of these problems, and relating that to their algorithmic tractability, has been made recently via their relations to spin glass models e.g.,~\cite{DMS15,DSS15,DSS16}. 

Following that analogy, the local minima of $\mathcal H(S)$ naturally correspond to the local MINCUTs of the complete graph  with edge-weights $J_{ij}$ (notice that in the case $\nu \sim \mbox{Ber}(p)$, this corresponds to local MINCUTs of the Erdos--Renyi random graph $\mathcal G(N,p)$). While extremal cuts are believed to be algorithmically intractable to find, even in random instances of the underlying graph, it is expected that locally extremal cuts are much easier to find, e.g.,~\cite{ABPW}. Indeed the local MAXCUT problem has seen a lot of attention and is believed to be algorithmically tractable for almost all random edge-weights. 

The corresponding local MINCUT problem we are analyzing is complicated by the fact that we want to avoid the homogeneous all-plus and all-minus solutions, and therefore it corresponds more precisely to a local version of the famous NAE (not all equal) constraint satisfaction problems. Based on the above heuristic connections and the success of our algorithms to find local minima even for large $N$, we propose the following.  

\begin{question}
Prove the existence of a polynomial time search algorithm for (non-trivial) local MINCUTs of $\mathcal G(N,p)$ for $p$ fixed (that succeeds with probability $1-o(1)$). More generally, prove the existence of such an algorithm for any non-degenerate $\nu$ supported on $\mathbb R_+$.  
\end{question}

\section*{Acknowledgments}
This work was supported in part through the NYU IT High Performance Computing resources, services, and staff expertise. The research of CMN and EYS was supported in part by US NSF grant DMS-1507019. DLS thanks the Aspen Center for Physics, supported by National Science Foundation grant PHY-1607611, where part of this work was performed.

\bibliographystyle{unsrt}
\bibliography{references}

\end{document}